\documentclass[aps,prd,nofootinbib]{revtex4}

\usepackage{amsmath}
\usepackage{amssymb}
\usepackage{graphicx}

\begin{document}

\title{The variation of $G$ in a negatively curved space-time}

\author{Jos\'e P. Mimoso}
\email{jpmimoso@cii.fc.ul.pt}\affiliation{Centro de Astronomia e
Astrof\'{\i}sica da Universidade de Lisboa, Campo Grande, Ed. C8
1749-016 Lisboa, Portugal}

\author{Francisco S. N. Lobo}
\email{flobo@cii.fc.ul.pt} \affiliation{Centro de Astronomia e
Astrof\'{\i}sica da Universidade de Lisboa, Campo Grande, Ed. C8
1749-016 Lisboa, Portugal}

\date{\today}

\begin{abstract}

Scalar-tensor (ST) gravity theories provide an
appropriate theoretical framework for the variation of Newton's
fundamental constant, conveyed by the dynamics of a scalar-field
non-minimally coupled to the space-time geometry. The experimental
scrutiny of scalar-tensor gravity theories has led to a detailed
analysis of their post-newtonian features, and is encapsulated
into the so-called parametrised post-newtonian formalism (PPN). Of
course this approach can only be applied whenever there is a
newtonian limit, and the latter is related to the GR solution that
is generalized by a given ST solution under consideration. This
procedure thus assumes two hypothesis: On the one hand, that there
should be a weak field limit of the GR solution; On the other hand
that the latter corresponds to the limit case of given ST
solution. In the present work we consider a ST solution with
negative spatial curvature. It generalizes a general relativistic
solution known as being of a degenerate class (A) for its unusual
properties. In particular, the GR solution does not exhibit the
usual weak field limit in the region where the gravitational field
is static. The absence of a weak field limit for the hyperbolic GR
solution means that such limit is also absent for comparison with
the ST solution, and thus one cannot barely apply the PPN
formalism. We therefore analyse the properties of the hyperbolic
ST solution, and discuss the question o defining a generalised
newtonian limit both for the GR solution and for the purpose of
contrasting it with the ST solution. This contributes a basic
framework to build up a parametrised pseudo-newtonian formalism
adequate to test ST negatively curved space-times.

\end{abstract}

\maketitle

\section{Introduction}
\label{sec:1} The possibility that physics might differ in diverse
epochs and/or places in the universe is a question of paramount
importance to understand  what are the limits of our present
physical
laws~\cite{Will:2005va,Damour:2002vu,Martins:2009zz,Barrow:2005hw}.
This issue is at present very much at the forefront of  the debate
in gravitational physics and cosmology\footnote{For various
perspectives on this issue see the other contributions in this
volume} as a result of the observations of a possible variation of
the fine structure constant $\alpha_{em}$ at high redshifts ($z>
0.5$) by Webb et al~\cite{Webb:2000mn}. These observations remind
us that our physics is based on peculiar coupling constants that
might also be evolutionary on the cosmological scale.

Variations of fundamental constants are a common feature in the
generalizations of Einstein's theory of general relativity
(GR)~\cite{Will:2005va}. Extensions of GR have not only been
claimed to be unavoidable when approaching the Planck scale of
energies, since gravitation is expected to be  unified with all
the other fundamental interactions, but they have also been
advocated  as an  explanation for the late time acceleration of
the universe recently unveiled by cosmological
observations~\cite{Lobo:2008sg,Bertolami:2007gv,Nunes:2009dj,Odintsov
2010}.

Scalar-tensor (ST) gravity theories, in particular, provide an
appropriate theoretical framework for the variation of Newton's
gravitational constant, which is induced by the dynamics of a
scalar-field non-minimally coupled to the space-time geometry. The
experimental scrutiny of scalar-tensor gravity theories requires a
detailed analysis of their post-newtonian features, and is
encapsulated into the so-called parametrised post-newtonian
formalism (PPN)
~\cite{Will:2005va,Damour:2002vu,Martins:2009zz,Bohmer:2009yx}.
This procedure assumes two hypothesis: On the one hand, that there
should be a weak field limit of the GR solution; On the other hand
that the latter corresponds to the limit case of a given ST
solution.

In the present work we investigate the impact of a hyperbolic
geometry on the possible variation of Newton's constant $G$. This
question has been somewhat overlooked in the past, and, as we will
show in the present work, raises a fundamental question regarding
the physical interpretation of the results. To address this issue
we derive a new  scalar-tensor solution with an hyperbolic
threading of the  spatial hypersurfaces\cite{Lobo:2009du}.  Our
solution extends a general relativistic solution known as being of
a degenerate class A2 for its unusual
properties\cite{Stephani:2003tm,Ehlers & Kundt 1962}. The latter
GR solution is characterised by a threading of the spatial
hypersurfaces by means of  pseudo-spheres instead of spheres. It
does not exhibit the usual weak field limit in the region where
the gravitational field is static, because the gravitational field
has a repulsive character.  This absence of a weak field limit for
the hyperbolic GR solution means that such limit is also absent
for comparison with the ST solution, and thus one cannot barely
apply the PPN formalism. To address the latter question, we
believe that one should look at the perturbations of the general
relativistic limit rather than of the absent newtonian weak field.
At least this enables us to assess the effects of the variation of
$G$.

\section{Scalar-tensor gravity theories}
\label{sec:2} In the Jordan-Fierz frame, scalar-tensor gravity
theories can be derived  from the action
\begin{equation}
S = \int\, \sqrt{-g}\,\left\{\left[\Phi R
-\frac{\omega{\Phi}}{\Phi}\, (\nabla(\Phi))^2 \right] + 16\Pi
G_N\, {\cal L}_m \right\} \label{ST_action}
\end{equation}
where $R$ is the usual Ricci curvature scalar of a spacetime
endowed with the metric $g_{ab}$,  $\Phi$ is a scalar field,
$\omega(\Phi)$ is a dimensionless coupling function, $U(\Phi)$ is
a cosmological potential for $\Phi$, and ${\cal L}_m$ represents
the Lagrangian for the matter fields\footnote{Alternatively we may
cast the action as $ L_\varphi=F(\varphi) R - \frac{1}{2}\, g^{ab}
\varphi_{,a} \varphi^{,b} + 2  U(\varphi)+ 16 \pi {\cal L}_m \;$
where the non-minimally coupled scalar field has a canonical
kinetic energy term.} (note that in this work we shall use units
that set $c=1$). Since $\Phi$ is a dynamical field, the trademark
of these theories is the variation of $G=\Phi^{-1}$  and the
archetypal theory is Brans-Dicke  theory in which $\omega(\Phi)$
is a constant~\cite{Brans:1961sx}.

In this frame the energy-momentum tensor of the matter
$T^{ab}=2/\sqrt{|g|}\,\delta S_m/\delta g_{ab} $ is conserved,
i.e., $\nabla_a T^{ab}=0$. This means that the matter test
particles follow the geodesics of the spacetime metrics, and the
scalar field feels the presence of matter and influences the
spacetime curvature, and hence the metric. Therefore the notorious
feature of this class of theories is the latter non-minimal
coupling between the scalar field and the spacetime geometry, in a
similar way  to that of the dilaton of string theory. Due to this
coupling, the gravitational physics
is governed by this interaction and the derivation of exact
solutions is considerably more difficult than in
GR\cite{Barrow:1994nx}.

This transpires perhaps in a more transparent way if we recast the
theory in the so-called Einstein frame by means of an appropriate
conformal transformation. Following Damour and Nordvedt's
notation~\cite{Damour:1993id}, we rescale the original metric
according to ($g_{ab}\to {\tilde{g}}_{ab} =  A^{-2}(\varphi) \,
g_{ab}$, where $A^{-2}(\varphi)=(\Phi/\Phi_{\ast})$ with
$\Phi_{\ast}=G^{-1}$ being a constant that we  take to be the
inverse of Newton's gravitational constant, and $\frac{{\rm
d}\ln{\Phi}}{{\rm d}\varphi} = \sqrt{\frac{16
\pi}{\Phi_\ast}}\,\alpha(\varphi)$). The action becomes
\begin{equation}
 {\cal L}_\varphi =\tilde R -
 \tilde{g}^{ab}
\varphi_{,a} \varphi^{,b} + 2  U(\varphi)+ 16 \pi \tilde{{\cal
L}_m}(\Psi_m,A^2(\varphi)\tilde{g}_{ab}) \;.
\end{equation}

Still as in Damour and Nordvedt~\cite{Damour:1993id} we introduce
\begin{equation}
{\cal{A}}(\varphi) = \ln A(\varphi) \quad , \qquad
\alpha(\varphi) =
\frac{\partial {\cal{A}}(\varphi)}{\partial \varphi}\quad , \qquad
{\cal{K}}(\varphi) = \frac{\partial \alpha(\varphi)}{\partial
\varphi} \; .
\end{equation}
Setting $U=0$, the field equations read
\begin{eqnarray}
\tilde{R}_{ab} &=& 2\partial_a \varphi\,\partial_b \varphi + 8\pi
G_N\,\left( \tilde{T}_{ab}-\frac{1}{2}\tilde{T}\,
\tilde{g}_{ab}\right) \\
\nabla^a\,\nabla_a\varphi &=& -4\pi\,G_N\,\tilde{T} \; .
\end{eqnarray}

This frame has the advantage of decoupling the helicities of the
linearised gravitational waves arising as metric perturbations
from the massless excitations of the scalar field $\varphi$.
Moreover, we can associate with the redefined scalar field the
role of a matter source acting on the right-hand side of the field
equations by introducing an adequate, effective energy-momentum
tensor. The net result can be interpreted as field equations in
the presence of two interacting sources: the redefined scalar
field and the original matter fields.  This mutual coupling
between the two components is dependent on  $\omega(\phi)$, and is
thus, in general, time
varying~\cite{Mimoso:1998dn,MimosoNunes03,Mimoso:1994wn}. The only
exception occurs when $\omega$ is constant, which corresponds to
the BD case. Different scalar-tensor theories correspond to
different couplings.

There are not many scalar-tensor solutions of negatively curved
universes in the literature, and thus it is of considerable
interest to derive and discuss a solution which to the best of our
knowledge is new, albeit a vacuum one\cite{O'Hanlon+Tupper 72}. In
what follows we address this question by first reviewing the
general relativistic solution.

\section{The general-relativistic vacuum solution
with pseudo-spherical symmetry}

We consider the metric given by
\begin{equation}
ds^2= -e^{\mu(r)}\,{\rm d}t^2+ e^{\lambda(r)} \,{\rm
d}r^2+r^2\,({\rm d}u^2+\sinh^2{u}\,{\rm d}v^2) \,,
\label{metric_constnc}
\end{equation}
where the usual 2$-d$ spheres are replaced by pseudo-spheres,
${\rm d}\sigma^2={\rm d}u^2+\sinh^2{u}\,{\rm d}v^2$, hence by
surfaces of negative, constant curvature. These are still surfaces
of revolution around an axis, and $v$ represents the corresponding
rotation angle. For the vacuum case we get
\begin{equation}
e^{\mu(r)}=e^{-\lambda(r)} =\left(\frac{2\mu}{r}-1\right) \; ,
\label{metric_constnc-antiScwarz}
\end{equation}
where $\mu$ is a constant \cite{Stephani:2003tm,Bonnor & Martins 1991}.
This vacuum solution is referred as degenerate solutions of class A
\cite{Stephani:2003tm}, and being an axisymmetric
solution it is a particular case of Weyl's
class of solutions\cite{Bonnor & Martins 1991},

We immediately see that the static solution holds for $r<2\mu$ and
that there is a coordinate singularity at $r=2\mu$ (note that
$|g|$ neither vanishes nor becomes $\infty$ at
$r=2\mu$)\cite{Anchordoqui:1995wa,Lobo:2009du}. This is the
complementary domain of the exterior Schwarzschild solution. In
our opinion this metric can be seen as an anti-Schwarzschild in
the same way the de Sitter model with negative curvature is an
anti-de Sitter model. In the region $r>2\mu$, likewise what
happens in the latter solution, the $g_{tt}$ and $g_{rr}$ metric
coefficients swap signs and the metric becomes cosmological.

Using pseudo-spherical coordinates
\begin{equation}
\{x = r\sinh u  \cos v,\, y =
r\sinh u  \sin v,\,z = r \cosh u,\,w = b(r)\}\; ,
\end{equation}
the spatial part of the metric (\ref{metric_constnc}) can be
related to the hyperboloid
\begin{equation}
w^2+x^2+y^2-z^2 = \left(\frac{b^2}{r^2} - 1\right)\, r^2 \,,
\end{equation}
embedded in a 4-dimensional flat space. We then have
\begin{equation}
{\rm d}w^2 +{\rm d}x^2+{\rm d}y^2-{\rm d}z^2 = \left((b'(r))^2 -
1\right)\,{\rm d}r^2+ r^2 \left( {\rm d}u^2+\sinh^2u\, {\rm d}v^2
\right) \;,
\end{equation}
where the prime stands for differentiation with respect to $r$,
and
\begin{eqnarray}
b(r)
= \mp 2\sqrt{2\mu}\sqrt{2\mu-r}\, .
\end{eqnarray}
It is possible to write the line element as
\begin{eqnarray}
{\rm d} s^2= -\, \tan^2\left[\ln\left( \bar r\right)^{\mp 1}
\right] \,{\rm d}\tau^2+
 \left(\frac{2\mu}{\bar r}\right)^2\,
 \cos ^4\left[\ln\left( \bar r\right)^{\mp 1}\right]\,\times
   \nonumber  \\
\times \left[{\rm d}\bar{r}^2\,+ \bar r^2\,({\rm d}u^2
+\sinh^2{u}\,{\rm d}\phi^2) \right]\; , \label{Isotrop3}
\end{eqnarray}
which is the analogue of the isotropic form of the Schwarzschild
solution. In the neighbourhood of $u=0$, i.e., for $u\ll 1$, we
can cast the metric of the 2-dimensional hyperbolic solid angle as
\begin{equation}
{\rm d}\sigma^2 \simeq  {\rm d}u^2 + u^2 {\rm d}v^2 \,
\end{equation}
so that it may be confused with the tangent space to the
spherically symmetric $S^2$ surfaces in the neighborhood of the
poles. The apparent arbitrariness of the locus $u=0$, is overcome
simply by transforming it to another location by means of a
hyperbolic rotation, as it is done in case of the spherically
symmetric case where the poles are defined up to a spherical
rotation (SO(3) group). So, the spatial surfaces are conformally
flat. However, we cannot recover the usual Newtonian weak-field
limit for large $r$, because of the change of signature that takes
place at $r=2\mu$.

In what concerns light rays, fixing $u$ and $v$ we have
\begin{equation}
\frac{{\rm d}t}{{\rm d}r} = \pm \frac{r}{2\mu-r}
\end{equation}
and we see that this ratio vanishes at $r=0$, becomes equal to one
at $r=\mu$ and diverges at $r=2\mu$. This tells us that, similarly
to the Schwarzschild solution, the light cones cones close
themselves when they approach the $r=2\mu$ event horizon, but
otherwise behave exactly in the opposite way to what happens in
the Schwarzschild exterior solution. Indeed the Schwarzschild's
outgoing light rays now become ingoing, and conversely.

Analysing the ``radial'' motion of test particles, we have the
following equation
\begin{equation}
\dot{r}^2+ \left(\frac{2\mu}{r} -1\right)\,\left(1+\frac{h^2} {r^2
\sinh^2 u_\ast } \right) = \epsilon^2\,,
\end{equation}
where $\epsilon$ and $h$ are constants of motion defined
by
$\epsilon= \left(\frac{2\mu}{r} -1\right)\,
\dot{t} = {\rm const_t}$
and
$h^2= r^2 \,\sinh^2 u_\ast \, \dot{v} = {\rm const_v}$,
for fixed $u=u_\ast$, and represent the energy and angular
momentum per unit mass, respectively.

\begin{figure}[ht]
\centering
\includegraphics[width=2.6in]{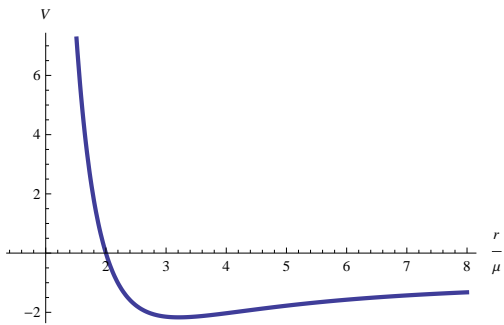}
\caption{\footnotesize Plot of the anti-Schwarzschild potential
$V(r/2\mu)$.} \label{figV}
\end{figure}

We may define the potential
\begin{equation}
2V(r) = \left(\frac{2\mu}{r} -1\right)\,\left(1+\frac{h^2} {r^2
\sinh^2 u_\ast } \right) \;,
\label{geod_potential}
\end{equation}
which we plot in Figure \ref{figV}. This potential is manifestly
repulsive, crosses the $r$-axis at $r=2\mu$, and asymptotes to the
negative value $V_\infty=-1$ as $r\to \infty$. It has a minimum at
$r_{\pm} =(h^2\mp \sqrt{h^4 - 12 \mu^2 h^2})/(2\mu)$,
provided the angular momentum per unit mass $h$ takes a high
enough value. However this minimum, when it exists, falls outside
the $r=2\mu$ divide. So we realise that a test particle is subject
to a repulsive potential and its radial coordinate is ever
increasing, inevitably crossing the event horizon at $r=2\mu$. (A
more complete discussion of the geodesics can be found
in\cite{Anchordoqui:1995wa}).
In \cite{Bonnor & Martins 1991} it is hinted that the
non-existence of a clear newtonian limit is related to the
existence of mass sources at $\infty$, but no definite conclusions
were drawn.

\section{The scalar-tensor solution}
In order to derive the scalar-tensor generalization of  the metric
(\ref{metric_constnc}), we apply a  theorem by Buchdahl
\cite{Buchdahl 1959} establishing the reciprocity  between any
static solution of Einstein's vacuum field equations and a
one-parameter family of solutions of Einstein's equations with a
(massless) scalar field.  The Einstein frame description of the
scalar-tensor gravity theories fits into the conditions of the
Buchdahl theorem. Indeed in this frame, after the conformal
transformation of the original metric, we
have GR plus a massless scalar field which is now coupled to the
matter fields. Therefore, in the absence of matter we can use Buchdahl's theorem
and we are able to derive the scalar-tensor generalisation of the
negatively  curved metric we have been considering. Given the
metric (\ref{metric_constnc}), we derive the corresponding
scalar-tensor solution
\begin{eqnarray}
{\rm d} s^2 &=& -\left(\frac{2\mu}{r}-1\right)^B\,{\rm d}t^2+
\left(\frac{2\mu}{r}-1\right)^{-B}\;{\rm d}r^2  \\ \nonumber
 & & \qquad  +\left(\frac{2\mu}{r}-1\right)^{1-B}\,r^2\,({\rm d}u^2
+\sinh^2{u}\,{\rm d}v^2) \, , \label{ST_sol_constnc} \\
\varphi(r) &=& \sqrt{\frac{C^2(2\omega+3)}{16\pi}\varphi_0}\,
\ln\left(\frac{2\mu}{r}-1\right)  \; , \label{ST_sol_constnc_phi}
\end{eqnarray}
where
\begin{equation}
C^2= \frac{1-B^2}{2\omega+3} \qquad -1 \le B \le 1 \;.
\end{equation}
This clearly reduces to our anti-Schwarzschild metric
(\ref{metric_constnc}) in the GR limit when $B=1$, and hence $C=0$
implying that $G=\Phi^{-1}$ is constant (we assume $(2\omega+3)>0$
throughout). On the other hand this also shows that the solution
has two branches corresponding to $C =\pm
\{(1-B^2)/(2\omega+3)\}^{1/2}$.

Notice that as pointed out by Agnese and La Camera
\cite{Agnese:1985xj}, the $r=2\mu$ limit is no longer just a
coordinate singularity, but rather a true singularity as it can be
seen from the analysis of the curvature invariants. In the
spherically symmetric case, Agnese and La Camera show that the
singularity at $r=2\mu$ has the topology of a point, and hence the
event horizon of the black hole shrinks to a point. In the
Einstein frame this happens because the energy density of the
scalar field diverges\cite{David Wands 1993}. In the case under
consideration the $r=2\mu$ condition now corresponds to the areal
radius of the pseudo-spheres,
$R=\left(\frac{2\mu}{r}-1\right)^{(1-B)/2}\,r$ becoming zero.

Reverting $\varphi=\int\sqrt{\Phi_0(2\omega+3)/(16\pi)}\,{\rm
d}\ln(\Phi/\Phi_0)$, and the conformal transformation,
$g_{ab}=(2\mu/r-1)^{-C}\,\tilde{g}_{ab}$, we can recast this
solution in the original frame in which the scalar-field is
coupled to the geometry and the content is vacuum, i.e. the Jordan
frame. We derive
\begin{eqnarray}
\Phi(r) &=& \Phi_0\,
\left(\frac{2\mu}{r}-1\right)^C  \; , \label{ST_sol_JF_constnc_phi} \\
{\rm d} s^2 &=& -\left(\frac{2\mu}{r}-1\right)^{B-C}\,{\rm d}t^2+
\left(\frac{2\mu}{r}-1\right)^{-B-C}\;{\rm d}r^2 \nonumber\\
 & & \qquad +\left(\frac{2\mu}{r}-1\right)^{1-B-C}\,r^2\,({\rm d}u^2
+\sinh^2{u}\,{\rm d}v^2) \, . \label{ST_sol_JF_constnc}
\end{eqnarray}
This shows that the gravitational constant $G =\Phi^{-1}$ decays
from an infinite value at $r=0$ to  a vanishing value at $r=2\mu$
when $C>0$, and conversely, grows from zero at $r=0$ to become
infinite at $r=2\mu$, when $C>0$. As we did for the general relativistic case we study the geodesic
behaviour of test particles in the scalar-tensor spacetime. The
quantities $\epsilon$ and $h$ defined as the energy per unit mass
and the angular momentum per unit mass, respectively, now become
\begin{eqnarray}
\epsilon = \left(\frac{2\mu}{r} -1\right)^{B-C}\,\dot{t}\;
\qquad , \qquad
h = \left(\frac{2\mu}{r} -1\right)^{1-B-C}\,\left(r^2 \,
\sinh^2 u_\ast \, \dot{v}\right)\; ,
\label{ang_mom_geod_ST}
\end{eqnarray}
and are once again first integrals of the motion of test
particles. Therefore for the ``radial'' motion of test particles,
we have the following equation
\begin{equation}
 \left(\frac{2\mu}{r} -1\right)^{-2C}\,\dot{r}^2+
 \left(\frac{2\mu}{r} -1\right)^{B-C}\,\left(1+\frac{h^2}
{\left(\frac{2\mu}{r} -1\right)^{(1-B-C)}\,r^2
\sinh^2 u_\ast } \right) = \epsilon^2 \; .
\label{STgeod_r_en_cons}
\end{equation}

\begin{figure} 
\includegraphics[width=2.6in]{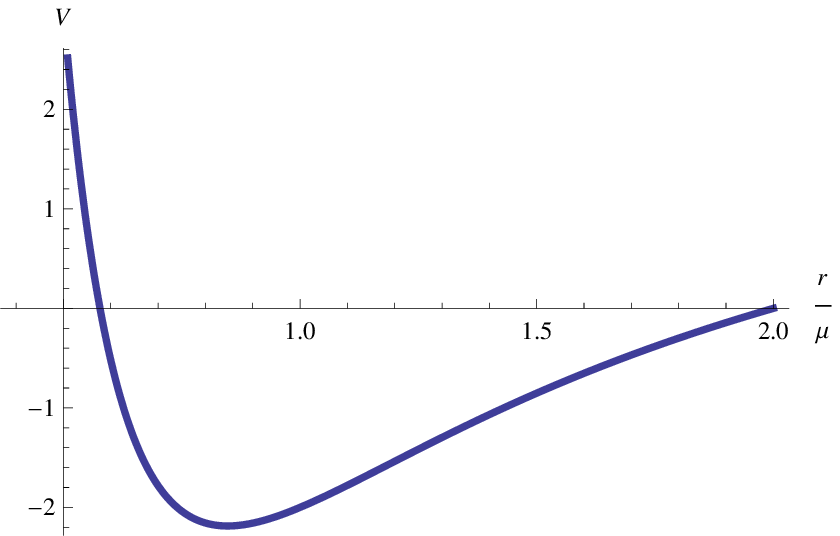}
\hspace{30pt}
\includegraphics[width=2.6in]{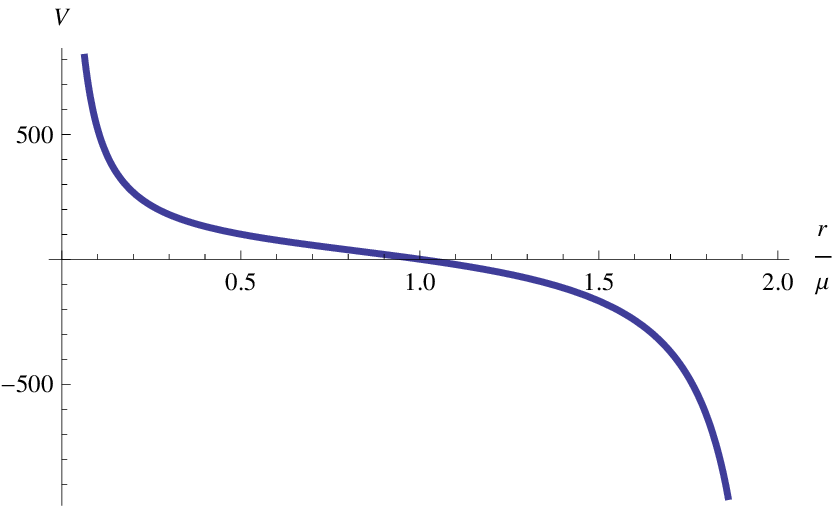}
\caption{In both plots $\omega=0$, $B=1/2$, $h_\ast=1$,
$\epsilon=2$, but left one has $C=+1/2$, and right $C=-1/2$.}
\label{Figure2}
\end{figure}

\begin{figure} 
\includegraphics[width=2.6in]{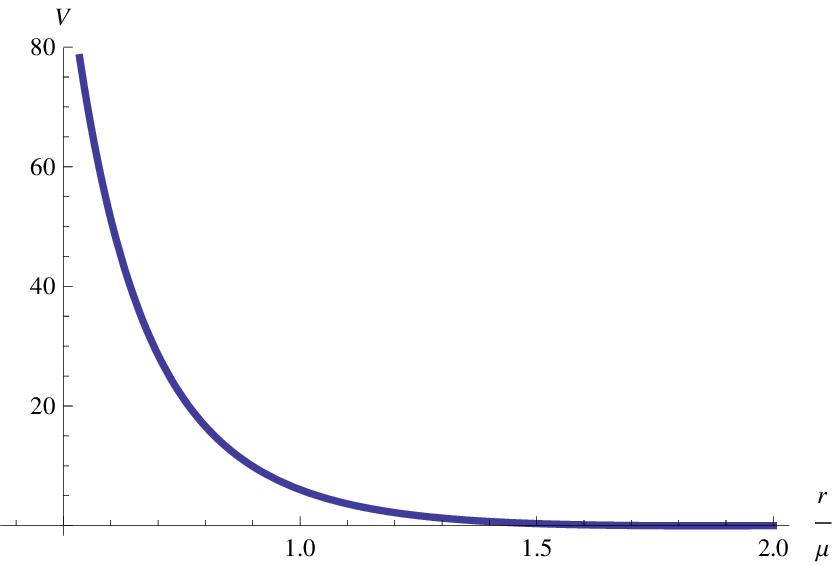}
\hspace{30pt}
\includegraphics[width=2.6in]{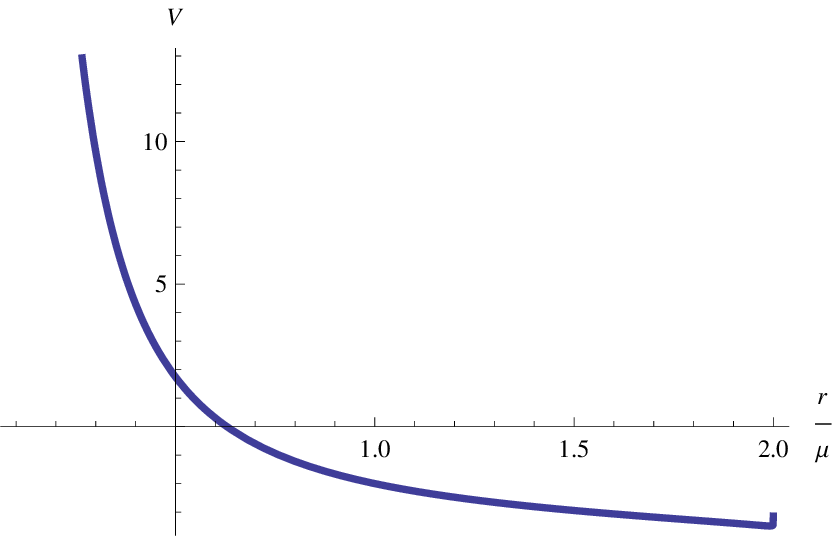}
\caption{Left plot: $\omega=0$, $B=1/2=C$,$\epsilon=2$, but
$h_\ast=2$, while right plot: $\omega=10^4$, $B=1/2$, $C=1/200$,
$\epsilon=2$, $h_\ast=1$} \label{Figure3}
\end{figure}

This is analogous to the equation of motion of a particle with
variable mass under the potential
$2V(r) = \left(\frac{2\mu}{r} -1\right)^{B-C}\,\left(1+\frac{h^2}
{\left(\frac{2\mu}{r} -1\right)^{(1-B-C)}\,r^2 \sinh^2 u_\ast }
\right) \;
$.
The latter crucially depend on the signs of the exponents of the
terms
$\xi(r) = \left(\frac{2\mu}{r} -1\right) \; .$
If we recast Eq. (\ref{STgeod_r_en_cons}) as
\begin{equation}
 \dot{r}^2+ \xi^{B+C}\,\left(1+\frac{h^2}
{\xi^{(1-B-C)}\,r^2 \sinh^2 u_\ast } \right) - \epsilon^2 \,
\xi^{2C} = 0\,\; ,
\label{STgeod_r_en_cons_B}
\end{equation}
we now have the motion of a test particle with vanishing effective
energy under the self-interaction potential
\begin{equation}
2V_{eff}(r) = \xi^{B+C}(r)\,\left(1+\frac{h_\ast^2}{r^2 }\,
\xi^{B+C-1}\right) - \epsilon^2\, \xi^{2C}\;,
\label{ST_geod_potential2}
\end{equation}
where $h_\ast=h\sinh u_\ast$. In Figures \ref{Figure2} and \ref{Figure3} we plot some possible
cases, which help us draw some important conclusions.  On the one
hand, high values of  $\omega$ imply more repulsive potentials,
since the higher $\omega$ the closer we are to GR. Notice that
high values of $\omega$ mean small $C$, i.e., smaller variation of
$G$. It is though remarkable that for larger departures from GR
(left plot of \ref{Figure2}) $V(r)$ may exhibit a minimum in the
range $0<r<2\mu$, leading to closed orbits, something which was
not possible in the GR solution. On the other hand  comparing the
left plots of Figures \ref{Figure2} and \ref{Figure3}, we realise
that the increase in angular momentum renders the potential more
repulsive, shifting the minimum beyond $r=2\mu$.

Overall, what is most remarkable in what regards the vacuum ST
solution derived here is that we are in the presence of a strong
gravitational field. The absence of a newtonian asymptotic limit
at the GR level is the signature of this situation, and prevents
us from performing the usual PPN multipolar expansion that permits
to identify the departures from GR.
Thus, if one wishes to ascertain how our ST solution departs from
GR, we need to look at the perturbation of the GR solution itself
(rather than that of the almost Minkowski weak field solution).
The way to accomplish this is to generalise the formalism
developed in a number of remarkable works for the Schwarzschild
solution (see \cite{Martel:2005ir} and references therein). We
have to trade the spherical symmetry of the latter by the
pseudo-spherical symmetry of our solution. At present we are
pursuing this task and we will report our  results elsewhere. From
the observational viewpoint what will be needed to test the
admissibility of the negatively curved solutions under
consideration (both the GR and the ST solutions) is to resort to
test of strong fields requiring the detection of gravitational
waves (for a discussion see \cite{Psaltis:2008bb})).


\section{Discussion}

We have considered a static solution with a pseudo-spherical
foliation of space. We reviewed its exotic features, and derived
the extended scalar-tensor solution. The fundamental feature of
these solutions is the absence of a newtonian weak field limit.
Indeed it is known that not all of the GR solutions allow a
newtonian limit, and this is the situation here. However, assuming
that the solutions of the Einstein field equations represent
gravitational fields, albeit far from our common physical
settings, it is possible to ascertain the implications of varying
$G$ in the strong fields by comparing the ST to their GR
counterparts. From the viewpoint of observations this relies on
the future detection of gravitational waves.
We conclude with a quotation from John Barrow
~\cite{Barrow:1991fj} which seems appropriate here
\begin{quote}
{\em The miracle of general relativity is that a purely
mathematical assembly of second-rank tensors should have anything
to do with Newtonian gravity in any limit}.

\end{quote}

\section*{Acknowledgement}
The authors are grateful to the organizers of the Symposium for a
very enjoyable atmosphere, and acknowledge the financial support
of the grants PTDC/FIS/102742/2008 and CERN/FP/109381/2009 from
FCT (Portugal).

\end{document}